\documentclass[aps,pra,twocolumn,groupedaddress]{revtex4-1}

\usepackage{graphicx}
\usepackage{soul}

\usepackage{natbib}

\begin{document}

\title{On the Uncertainty of the Ordering of Nonlocal Wavefunction Collapse when Relativity is Considered}

\title{On the Uncertainty of the Ordering of Nonlocal Wavefunction Collapse when Relativity is Considered}
\author{Chris D. Richardson$^1$ and Jon P. Dowling$^2$}
\affiliation{$^1$Department de Physique, University of Liege}
\affiliation{$^2$Department of Physics and Astronomy, Louisiana State University}
\date{\today}							

\begin{abstract}
The temporal measurement order and therefore the originator of the instantaneous collapse of the wavefunction of a spatiality entangled particle pair can change depending on the reference frame of an observer.  This can lead to a paradox in which its seems that both measurements collapsed the wavefunction before the other.  We resolve this paradox by demonstrating how attempting to determine the order of measurement of the entangled pair introduces uncertainty which makes the measurement order impossible to know.
\end{abstract}

\maketitle

\section{Introduction\label{sec:intro}}

The quantum mechanical instantaneous collapse of the wavefunction of distantly separated entangled particles has been a thorn in the side of relativity ever since its inception.  We of course know that causality is safe and no information can travel faster then the speed of light but the instantaneous and irreversible nature of wavefunction collapse can still lead to problems with relativity and quantum mechanics even though they are separately and together extremely well proven theories.  In particular it leads to a paradox, first found and proposed by Suarez and Scarani~\cite{bib:scarani}, in which two observers disagree about who collapsed the wavefunction of an entangled particle pair first.

Two particles are entangled if properties of the particles are correlated but indeterminate until a measurement is made.  Until that measurement is made the entangled particles share one quantum state, one inseparable wavefunction.   When a measurement is made on one of the entangled pair the wavefunction describing the entangled state collapses instantaneously and irreversibly into definite and separate correlated states.  For a pair of spatially entangled particles the collapse occurs for both photons instantaneously across the entire distance between them.  If projective measurements, which cause the collapse, are performed on both particles it is reasonable to ask which measurement was responsible for the collapse of the wavefunction and expect a definite answer.  However, according to relativity the timing of those measurements are relative to the reference frame of the observer.  The observers can disagree on which measurement collapsed the wavefunction and, of interest here, both observers can think that their measurements were the cause of the collapse.  This disagreement is the basis of the aforementioned paradox.

This paradox has led to many experiments~\cite{bib:wheel,bib:wheel2,bib:wheel3,bib:wheel4} and while they produce results consistent with both relativity and quantum mechanics they do not directly address the question of which measurement was responsible for the collapse of the wavefunction.  We will resolve this issue and use the results to speculate on the meaning of simultaneity within quantum mechanics.  To do so we first devise a thought experiment in section~\ref{sec:texp} that tests the measurement order of an entangled pair with detectors and observers that disagree about their measurement order.  We show in section~\ref{sec:constraint} that the experiment constrains the system in such a way that an uncertainty is induced into the system.  We use this constraint to construct the initial wavefunction in section~\ref{sec:initial} and in section~\ref{sec:transform} we observe how the wave function transforms when a projective measurement is made by the moving observer.  In section~\ref{sec:uncertainty} we find that the uncertainty inherent in the experiment makes the question of who measured first unanswerable and thus resolves the paradox.  First we will review the paradox in more detail.

\section{The Paradox\label{sec:paradox}}

\begin{figure}[pb]
  \includegraphics[width=1\columnwidth]{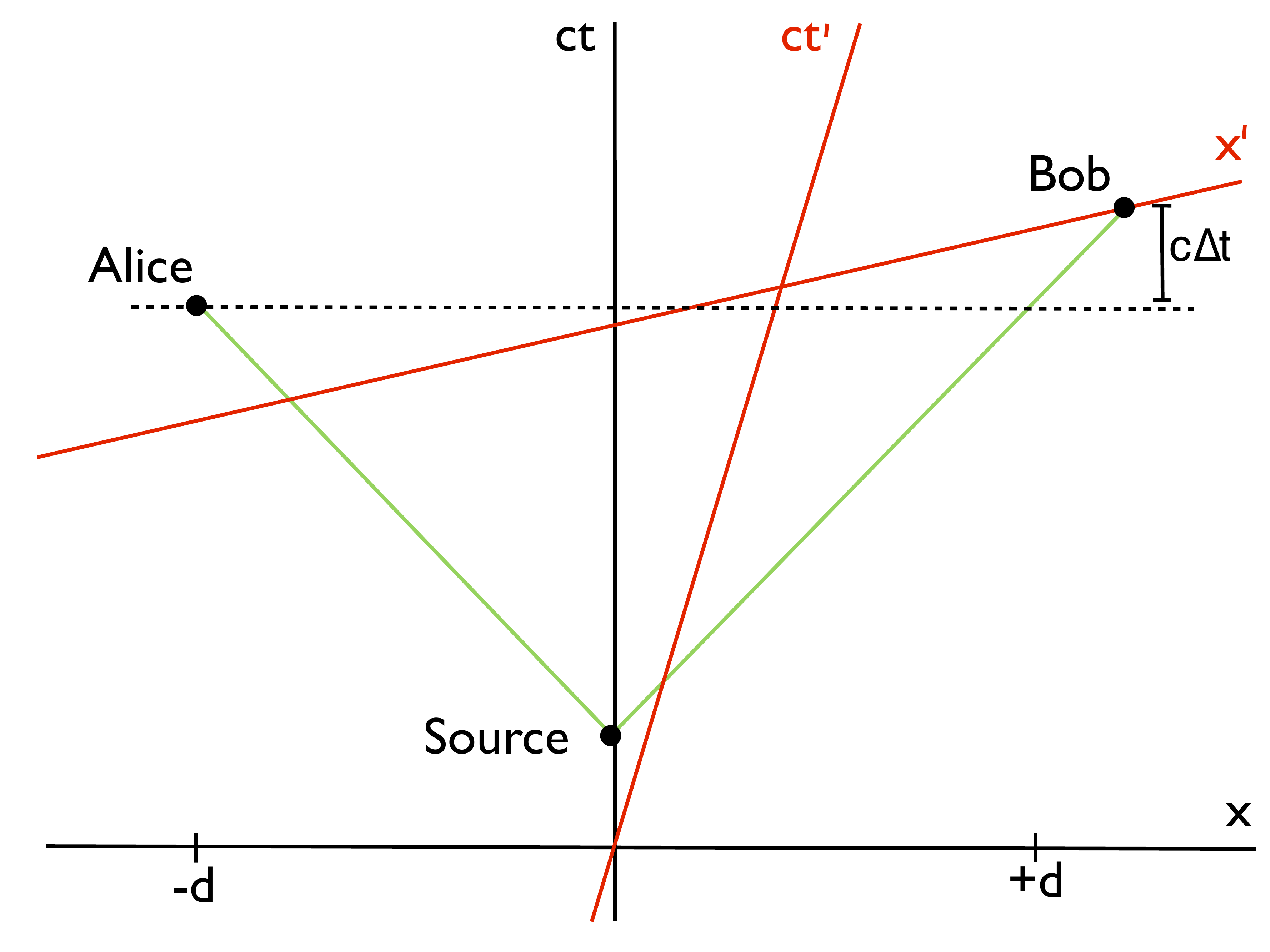}
\caption{A time-energy entangled pair is emitted and travels to Alice and Bob who is further from the source and in motion relative to Alice.  In the situation pictured both Alice and Bob measure first and they are said to be in a state of paradox.}
\label{fig:Zbinden_Wheel_ST_Diagram}
\end{figure}

As shown in Fig.~(\ref{fig:Zbinden_Wheel_ST_Diagram}) , two measuring devices that detect the time of arrival of a time-energy entangled biphoton~\cite{bib:biphoton}  are separated by a distance $2 d$.  One detector is further from the source and thus there is a time difference between the measurement events $\Delta t$.  Under a Lorentz transform this difference transforms as
\begin{equation}
	\Delta t' = \gamma (\Delta t - \frac{\beta (2 d)}{c}) \;, \label{eqn:origtransform}
\end{equation}
where $\beta = v/c$ and $\gamma = 1/\sqrt{1-\beta^2}$.  If the situation occurs in which
\begin{equation}
	\frac{\beta (2 d)}{c} > \Delta t\;, \label{eqn:ineq}
\end{equation}
then $\Delta t'$ becomes negative and the time ordering of the measurements reverse.  Alice, an observer in the lab frame, and Bob, and observer in the frame in which Eq. (\ref{eqn:ineq}) is true, will disagree about which measurement was responsible for the collapse of the wavefunction and we will now refer to them as being in a \emph{state of paradox}.

We need to be careful here about the difference between an observer in the context of relativity and a measurement in context of quantum mechanics.  In relativity an observer does not necessarily need to be the one making measurement.  This disconnect between observer and measurement is incompatible with quantum mechanics in which an observation is equivalent to a measurement.  To be consistent between the theories we must insist that all observers are \emph{measuring observers}.  We will say that Alice performs a measurement on one half of the biphoton in the lab frame and Bob performs the measurement on the other half of the biphoton in the moving frame.  We will next devise a thought experiment that attempts to determine the temporal order of these two measurements when the two measuring observers, Alice and Bob, are in a state of paradox.

\section{The Thought Experiment\label{sec:texp}}
 
\begin{figure*}[ht]
	\begin{minipage}[t]{0.95\columnwidth}
 \centering
 \includegraphics[width=1\columnwidth]{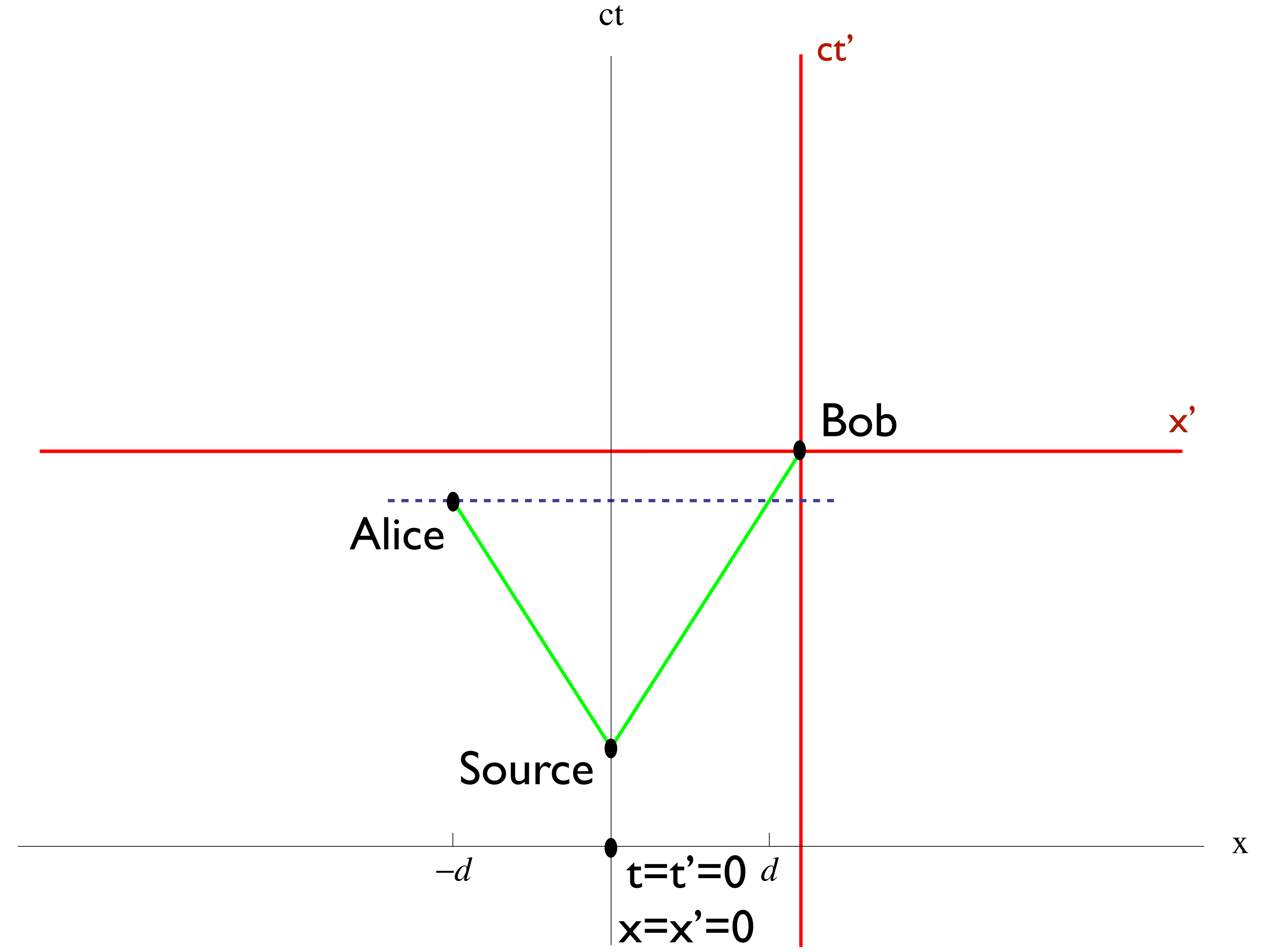}
\caption{Bob is stationary.  At some time $t_e$ after Bob passes the biphoton source, a biphoton is emitted.  In both Alice and Bob's frame, Bob detects a photon after Alice.}
\label{fig:SpaceTimeNoVelocity}
 \end{minipage}
 \hspace{0.5cm}
 \begin{minipage}[t]{0.95\columnwidth}
 \centering
 \includegraphics[width=1\columnwidth]{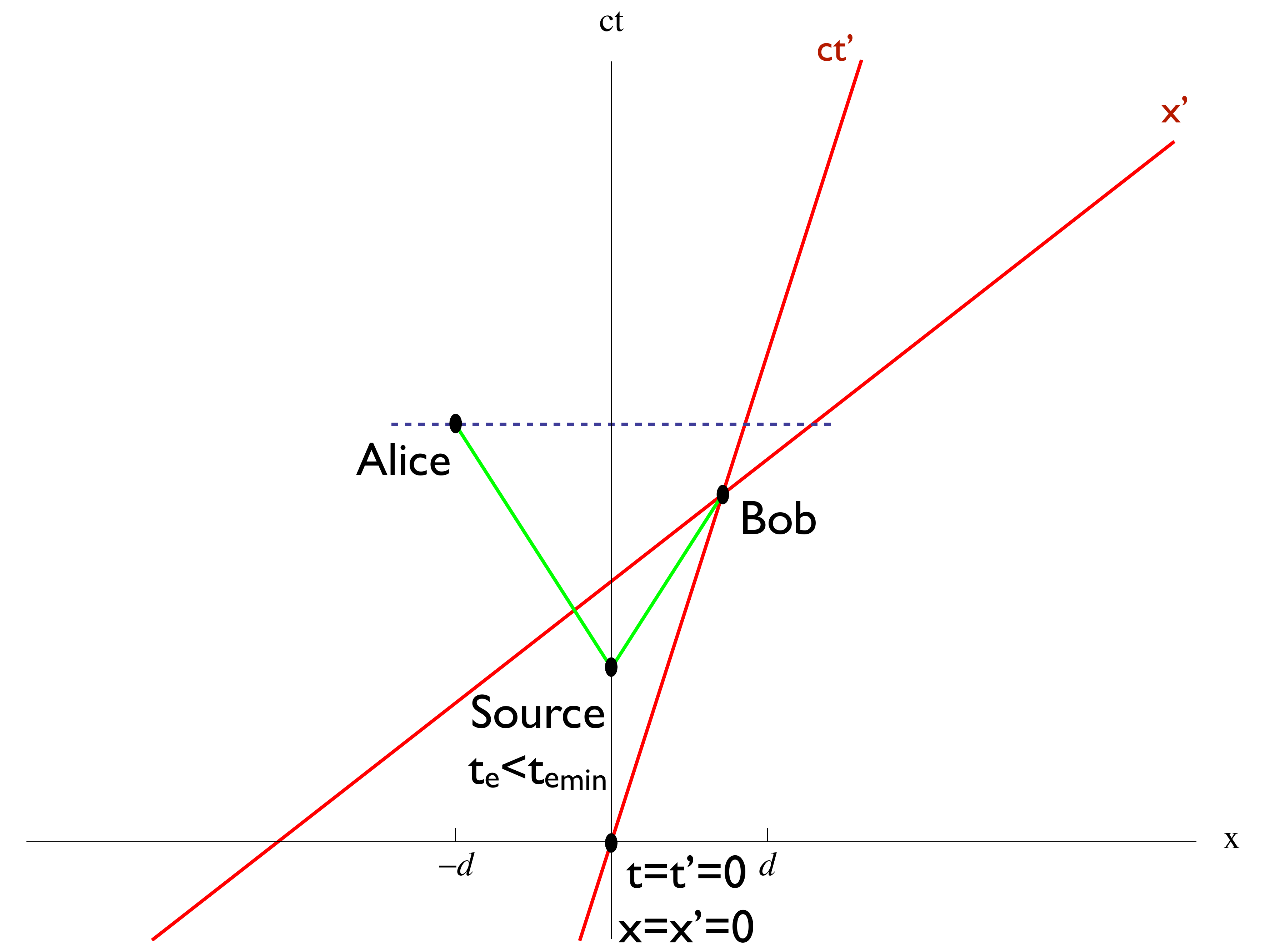}
\caption{At some time $t_e$ where $t_e < t_{e_{min}} =  d (1/\beta - 1) / c$ after Bob passes the biphoton source, a biphoton is emitted.  In both Alice and Bob's frame, Bob detects a photon before Alice.}
\label{fig:SpaceTimeEarly}
 \end{minipage}
 \hspace{0.5cm}
 \begin{minipage}[t]{0.95\columnwidth}
 \centering
 \includegraphics[width=1\columnwidth]{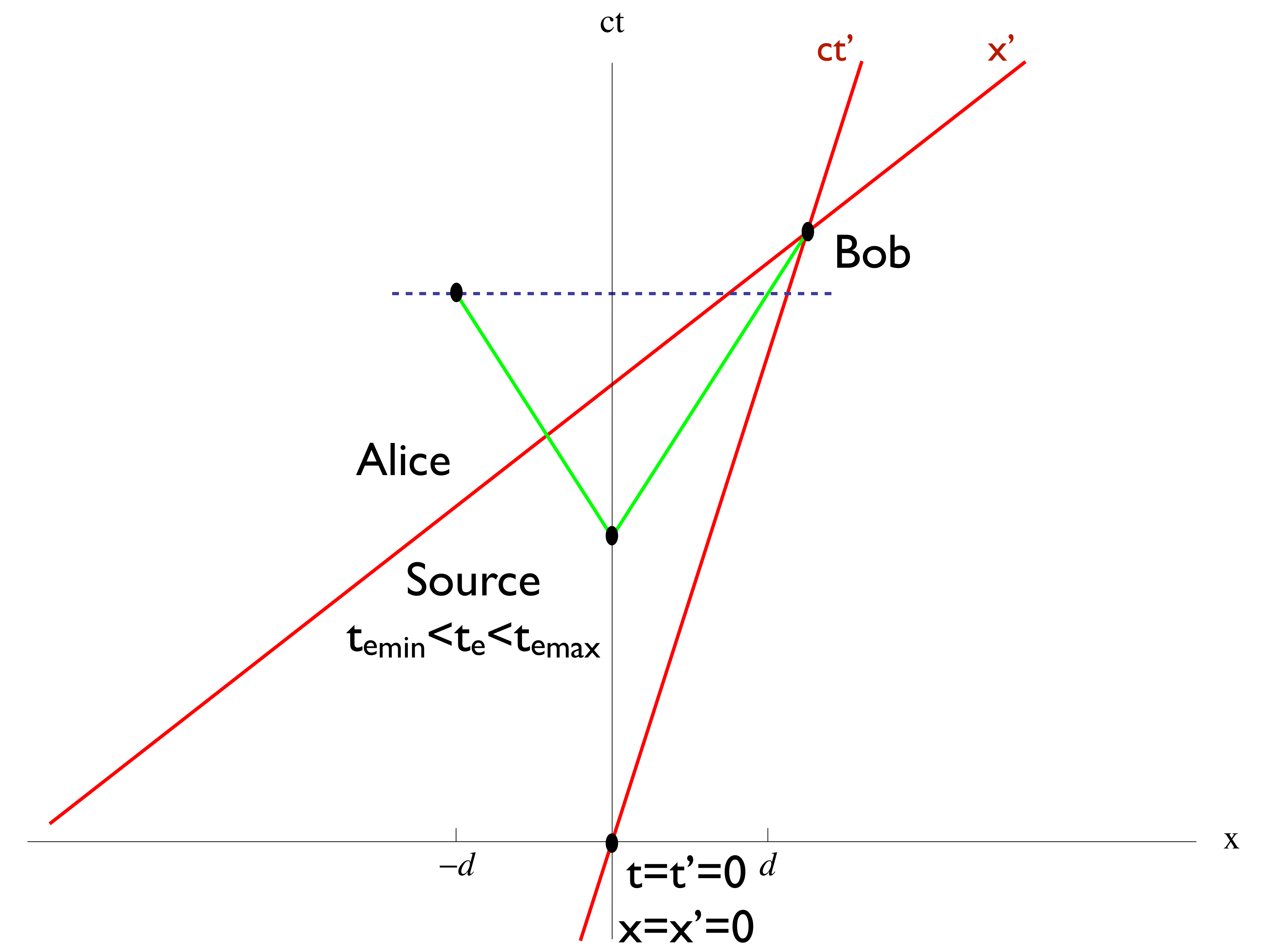}
\caption{At some time $t_e$ where $d (1/\beta - 1) / c = t_{e_{min}} < t_e < t_{e_{max}} =  d (1/\beta + 1) / c$ after Bob passes the biphoton source, a biphoton is emitted.  In Alice's reference frame, the lab frame, Bob detects a photon after Alice.  In Bob's reference frame the measurement order is reversed.}
\label{fig:SpaceTime}
 \end{minipage}
 \hspace{0.5cm}
 \begin{minipage}[t]{0.95\columnwidth}
 \centering
 \includegraphics[width=1\columnwidth]{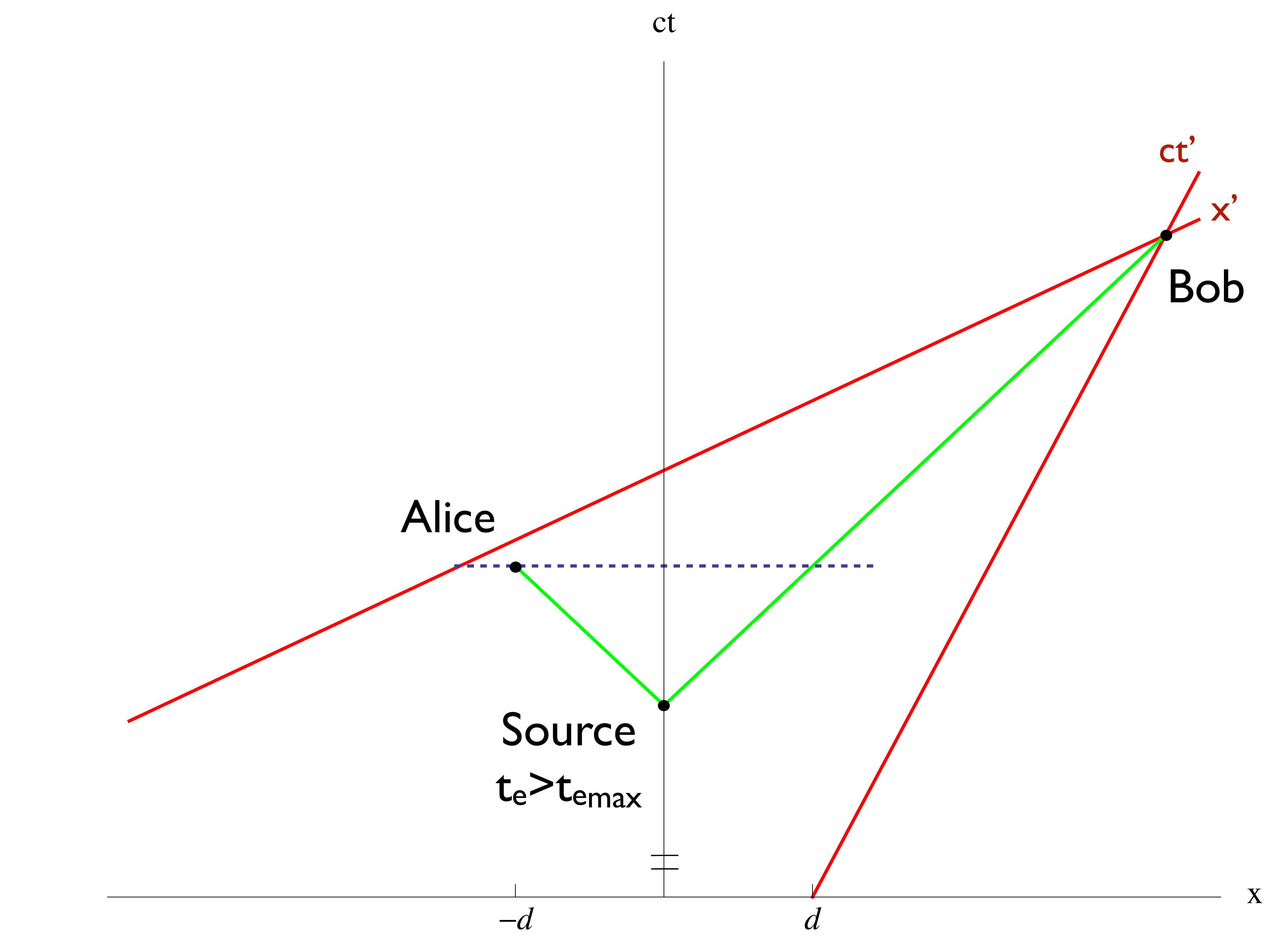}
\caption{At some time $t_e$ where $t_e > t_{e_{max}} =  d (1/\beta + 1) / c$ after Bob passes the biphoton source, a biphoton is emitted.  In both Alice and Bob's frame, Bob detects a photon after Alice.}
\label{fig:SpaceTimeBust}
 \end{minipage}
 \end{figure*}

The thought experiment is represented in the space-time diagrams in Figs.~(\ref{fig:SpaceTimeNoVelocity} - \ref{fig:SpaceTimeBust}).  Bob is in motion relative to Alice with a velocity $\beta$.  At $(x_{B_0},t_{B_0}) = (x'_{B_0},t'_{B_0}) = (0,0)$ Bob passes a source of time-energy entangled photons and Alice is at $(x_{A_0},t_{A_0}) = (-d,0)$.  At some later time $t_e$ the source emits a biphoton.  Alice measures her photon at $(x_A,t_A) = (-d, t_e + d/c)$ and Bob measures his photon at $(x_B,t_B) = (\beta \, c \, t_B, t_B)$.  It is assumed that Alice sends Bob a signal when her detector triggers and Bob can use it to determine Alice's measurement time and determine the difference in time $\Delta t$, which has a Lorentz transform of
\begin{eqnarray}
	\Delta t' &=&  \gamma (\Delta t - \frac{\beta}{c} \Delta x) = \gamma (\Delta t - \frac{\beta}{c} (2 d + \beta c \Delta t)) \;, \label{eqn:lorentzboostwithx} \\
	 &=& \gamma (\Delta t (1-\beta^2)-2 \frac{\beta}{c} d) \;.\label{eqn:lorentzboost}
\end{eqnarray}
As in Eq. (\ref{eqn:origtransform}) if the velocity, $\beta$, or the distance, $d$, between Alice and Bob gets large enough the temporal order perceived by Bob from Eq. (\ref{eqn:lorentzboost}) changes.  Therefore, for the experiment to be in the state of paradox $\beta$ or $d$ must be sufficiently large.  So far this experiment does not guarantee that the observers will be in a state of paradox.  To do that we must add a constraint to the emission time of the biphoton.

\section{The Emission Time Constraint\label{sec:constraint}}

It is necessary to realize that to test the situation in which the paradox occurs we \emph{must} require that the emission time of the biphoton to be constrained to a time period, $\Delta t_e$, during which it is possible for the paradox to exist.  If they are emitted outside this region like in Figs.~(\ref{fig:SpaceTimeEarly}) and (\ref{fig:SpaceTimeBust}), there can be no result since both observers will agree on the time ordering and hence no paradox.  The emission time constraint can be derived with some simple algebra.  For Bob to be in a position to make a measurement after Alice in Alice's frame, he must at least be equidistant from the source, which puts Alice and Bob's minimum position and time at
\begin{eqnarray}
	x_{A_{min}} &=& x_{B_{min}} = d  \;, \\
	t_{A_{min}} &=& t_{B_{min}} = \frac{d}{\beta c} \;.
\end{eqnarray}
When Bob leaves the maximum time in which Alice believes she measured first we can make a line from Alice to Bob that has the slope of Bob's velocity $t_{B_{max}} = \beta x_{B_{max}}/c + d (1+\beta)/c + t_{e_{max}}$.  When we set this equal to the line describing the photons path $t_{B_{max}} = t_{e_{max}} +  x_{B_{max}}/c$, we can get Alice and Bob's maximum position and time
\begin{eqnarray}
	x_{A_{max}} &=& d \;, \\
	t_{A_{max}} &=& \frac{d}{\beta c}(2 \beta + 1) \;, \\
	x_{B_{max}} &=& d \frac{1+\beta}{1-\beta} \;, \\
	t_{B_{max}} &=& \frac{x_{B_{max}}}{\beta c} = \frac{d}{\beta c} \frac{1+\beta}{1-\beta} \;. 
\end{eqnarray} 
To get the constraint on the emission time $\Delta t_e$ we take Bob's measurement times and subtract the amount of time it took the photon to get to Bob.
\begin{eqnarray}
	t_{e_{min}} &=& t_{B_{min}} -\frac{x_{B_{min}}}{c} = \frac{d}{\beta c}(1 - \beta)\;, \\
	t_{e_{max}} &=& t_{B_{max}} - \frac{x_{B_{max}}}{c} = \frac{d}{\beta c}(1 + \beta) \;.
\end{eqnarray} 
Therefore, the time constraint $\Delta t_e$ is
\begin{equation}
	\Delta t_e = t_{e_{max}} - t_{e_{min}} = \frac{2 d}{c} \;. \label{eqn:constraint}
\end{equation}

Simply posing the thought experiment in such a way that the paradox is ensured adds this time constraint to the emission time of the biphoton.  The experimental configuration gives us information about the entangled pair and any knowledge about the biphoton is considered a measurement on the biphoton.  We will show that this will lead to an uncertainty which balances the time difference induced by the Lorentz transform in Eq.~(\ref{eqn:lorentzboost}).

\section{The Initial Wavefunction\label{sec:initial}}

The paradox being discussed here hinges on the ability to measure the arrival time of a photon.  This can be problematic as there is no actual time operator and therefore no eigenstates of time of which to construct the wavefunction.  If, however, we deal only with photons we can follow Shalm \emph{et al.}\cite{bib:threephotonentanglement} and relate the time and energy operators and eigenstates to those of position and momentum which are well defined.  Photons are massless and travel at the speed of light and therefore measuring the arrival time is equivalent to a measurement of its position $x(t = x/c)$ and measuring its energy is equivalent to a measurement of its momentum $p(\hbar \omega = c p)$.  We can also therefore relate the eigenstates $| t \rangle = \frac{1}{c} | x \rangle$ and $| E \rangle = c | p \rangle$ and derive the relationships $\langle E | t \rangle =  e^{i E t}$, $\langle t | E \rangle =  e^{-i E t}$, $\langle t_1 | t_2 \rangle = \delta(t_1-t_2)$ and $\langle E_1 | E_2 \rangle = \delta(E_1-E_2)$.

We can now construct the initial wavefunction by starting with the following Einstein, Podolsky, Rosen (EPR)~\cite{bib:epr} states in the energy-time representation
\begin{eqnarray}
	\psi_0(t_A,t_B) &=& \langle t_A,t_B | \Psi_0  \rangle = \delta(t_A - t_B) \;, \\
	&=& \int_{-\infty}^{\infty} e^{- i E (t_A - t_B)} \,\mathrm{d}E  \;, \\
	\phi_0(E_A,E_B) &=& \langle E_A,E_B | \Psi_0 \rangle \;,\quad\quad\quad\quad\quad\quad\quad\quad\quad\quad\quad
\end{eqnarray}
\begin{eqnarray}
	\;\;\;\;&=&  \int_{-\infty}^{\infty} \mathrm{d}t_A\,  \int_{-\infty}^{\infty} \mathrm{d}t_B\, \langle E_A,E_B | t_A, t_B \rangle \langle t_A,t_B | \Psi \rangle , \\
	&=& \int_{-\infty}^{\infty} \mathrm{d}t_A \, \int_{-\infty}^{\infty} \mathrm{d}t_B \, e^{- i (E_A t_A - E_B t_B)} \delta(t_A - t_B) , \\
	&=& \int_{-\infty}^{\infty} \mathrm{d}t_A \, e^{- i t_A (E_A + E_B)} = \delta(E_A + E_B) \;, \label{eqn:epr_E}
\end{eqnarray}
where we are now using natural units with $c=\hbar=1$.  The time constraint from Eq. (\ref{eqn:constraint}) when applied to Eq. (\ref{eqn:epr_E}) will yield the constrained initial wavefunction
\begin{eqnarray}
	| \Psi_C  \rangle &=& \int_{-\infty}^{\infty}\mathrm{d}t_A \int_{-\infty}^{\infty}\mathrm{d}t_B  \Pi(t_A) | t_A,t_B \rangle \langle t_A,t_B | \Psi_0  \rangle , 
\end{eqnarray}
\begin{eqnarray}
	\psi_C(t_A,t_B) &=& \langle t_A,t_B | \Psi_C  \rangle = \delta(t_A - t_B) \Pi(t_A)\;, 
\end{eqnarray}
where the constraining function $\Pi(t_A)$ is
\begin{eqnarray}
	\Pi(t_A) &=& \left\{ 
  \begin{array}{l l}
    1/\sqrt{2d} & \quad \mathrm{if} \; d < t_A < 3 d\\
    0 & \quad \mathrm{otherwise}
  \end{array} \right.\;.
\end{eqnarray}
The wavefunction in the energy representation is
\begin{eqnarray}
	\phi_{C}(E_A,E_B) &=& \langle E_A,E_B | \Psi_C \rangle \;, \quad\quad\quad\quad\quad\quad\quad\quad\quad\quad\quad
\end{eqnarray}
\begin{eqnarray}
	&=& \int_{-\infty}^{\infty}\,\mathrm{d}t_A \int_{-\infty}^{\infty} \mathrm{d}t_B \; \langle E_A,E_B | t_a, t_B \rangle \langle t_A,t_B | \Psi_C \rangle  , \\
	&=&  \int_{-\infty}^{\infty}\,\mathrm{d}t_A \int_{-\infty}^{\infty} \mathrm{d}t_B \; e^{- i t_A E_A - i t_B E_B} \nonumber \\
	&&\times  \delta(t_A - t_B)  \Pi(t_A) \;, \\
	&=&  \int_{-\infty}^{\infty}\,\mathrm{d}t_A \; e^{- i t_A (E_A + E_B)}  \Pi(t_A) \;, \\
	&=& \frac{1}{\sqrt{\pi d}} \frac{\sin(d(E_A+E_B))}{(E_A+E_B)} e^{- i 2 d(E_A + E_B)} \;.
\end{eqnarray}
This wavefunction, with an obvious energy uncertainty, would not usually make a difference to a arrival time measurement since if Alice and Bob were in the same reference frame the energy component of the wave function would not be measured and the arrival time uncertainty would be zero.  This is not the case when we take into account Bob's measurement made from a moving frame.

\section{The Wavefunction Taking Into Account Bob's Measurement\label{sec:transform}}

The laws of physics are of course the same in all reference frames, but this does not mean that measurements made from different reference frames will yield the same results.  This is obvious if we consider the simple case of the Doppler shift.  In this thought experiment Alice and Bob are in difference frames, with relativistically scaled clocks and rulers,  and so the effects and results of their respective projective measurements will be different and must be taken into account.

In Bob's reference frame he uses his clock to make a projective measurement of the arrival time of the biphoton.  Assume that the projection has a vanishingly small uncertainty of $\delta t'_B$ which can be interpreted as the time it takes for Bob's measurement to be made.  Alice observes Bob's clock ticking slower than her own and therefore the amount of time that passes in Alice's reference during time in which Bob's measurement takes place is
\begin{eqnarray}
\delta t_B = \gamma \delta t'_B \;. \label{eqn:bobsuncertainty}
\end{eqnarray}
As Bob approaches the speed of light it looks to Alice as if the time it takes for Bob to make his measurement increases to infinity.  While Bob projects his half of the wavefunction to a function with a small uncertainty in his reference frame the projection when observed from Alice's moving frame looks to have a larger uncertainty given by Eq. (\ref{eqn:bobsuncertainty}).

If Bob is at rest relative to Alice his measurement would, unsurprisingly, project his half of the wavefunction into a time eigenstate and the wavefunction would be
\begin{eqnarray}
	\psi_{\beta=0}(t_A,t_B) &=&  \int_{-\infty}^{\infty} dt_A \int_{-\infty}^{\infty} dt_B \; \langle t_A, t_B | \Psi  \rangle \nonumber  \\
	&&\times \langle t_A,t_B | \delta(t_B-t_0) | t_A, t_B \rangle  \;, \\
	&=&  \delta(t_A-t_B) \delta(t_B-t_0) \;. \label{eqn:betazero}
\end{eqnarray}
However, as Bob's velocity reaches the speed of light he projects his part of the wavefunction onto a state which, according to Eq. (\ref{eqn:bobsuncertainty}), has an infinite time uncertainty, or in other words an energy eigenstate, and the wavefunction would be
\begin{eqnarray}
	\psi_{\beta=1}(t_A,t_B) &=&  \int_{-\infty}^{\infty} dE_A \int_{-\infty}^{\infty} dE_B \;  \langle E_A, E_B | \Psi  \rangle \nonumber  \\
	&&\times \langle t_A,t_B | \delta(E_B-E_0) | E_A, E_B \rangle \;, \\
	&=& \int_{-\infty}^{\infty} dE_A \, \frac{1}{\sqrt{\pi d}} \frac{\sin(2 d(E_A + E_0))}{(E_A + E_0)} \nonumber \\
	&&\times e^{i (E_A t_A + E_0 t_B)} \;, \\
	&=&  \Pi(t_A) e^{i E_0(t_A - t_B)} \;. \label{eqn:betaone}
\end{eqnarray}
We can construct a superposition of these two states which will describe the wavefunction for all the intermediate values for Bob's velocity
\begin{eqnarray}
 	| \Psi_{\beta} \rangle &=& A\; | \Psi_{\beta=0} \rangle   + B\; | \Psi_{\beta=1} \rangle \;, \\
	 \psi_{\beta}(t_A,t_B) &=& A\; \psi_{\beta=0}(t_A,t_B)   + B \; \psi_{\beta=1}(t_A,t_B) \;, \\
	&=& A\; \delta(t_A-t_B) \delta(t_B-t_0) \nonumber \\
	&+& B\; \Pi(t_A) e^{i E_0(t_A - t_B)} \;.
\end{eqnarray}
We can deduce $A$ by considering the wavefunction in Bob's reference frame.  We know that in Bob's frame he is in an eigenstate of time so
\begin{eqnarray}
	 \psi_{\beta}(t'_A,t'_B) &=& A\; \delta(t'_A-t'_B) \delta(t'_B-t'_0) \;, \\
	\psi_{\beta}(t_A,t_B) &=& A\; \delta((t_A-t_B)/\gamma) \delta((t_B-t_0)/\gamma) \;, \\
	&=& \gamma^2 A \; \delta(t_A-t_B) \delta(t_B-t_0) \;, 
\end{eqnarray}
and therefore $A =1/\gamma^2 = 1-\beta^2$ and consequently $B = \beta^2$.  The wavefucntion after Bob's measurement is therefore
\begin{eqnarray}
 	| \Psi_{\beta} \rangle &=&(1-\beta^2)| \Psi_{\beta=0} \rangle   + \beta^2 | \Psi_{\beta=1} \rangle \;, \\
	 \psi_{\beta}(t_A,t_B) &=& \langle t_A,t_B | \Psi_{\beta} \rangle \;, \\
	&=& (1-\beta^2)\psi_{\beta=0}(t_A,t_B) \nonumber \\
	&+& \beta^2 \, \psi_{\beta=1}(t_A,t_B) \;, \\
	&=& (1-\beta^2) \delta(t_A-t_B) \delta(t_B-t_0) \nonumber \\
	&+& \beta^2 \, \Pi(t_A) e^{i E_0(t_A - t_B)} \;, \label{eqn:pow}
\end{eqnarray}
noting $\langle  \Psi_{\beta} | \Psi_{\beta} \rangle = \int_{-\infty}^{\infty} dt_A \int_{-\infty}^{\infty} dt_B | \langle t_A,t_B | \Psi_{\beta} \rangle |^2 = 1$.

\section{The Uncertain Paradox\label{sec:uncertainty}}

After a projective measurement into Bob's reference frame Alice's time uncertainty, $\sigma_{t}$, of the wavefunction given by Eq. (\ref{eqn:pow}), is
\begin{eqnarray}
	\sigma^2_{t} = \beta^2 (2 d)^2 \;. \label{eqn:theuncertainty}
\end{eqnarray}
We can check the asymptotic behavior and see that as $\beta \rightarrow 1$ the time uncertainty vanishes and as $\beta \rightarrow 1$ the time uncertainty becomes the original time constraint, $\sigma_{t} = 2 d$ which is exactly what we would get from the wave functions in Eqs. (\ref{eqn:betazero}) and (\ref{eqn:betaone}).  Adding the uncertainty of Eq.~(\ref{eqn:theuncertainty}) to the boosted time difference in Eq.~(\ref{eqn:lorentzboostwithx}) (in natural units) gives
\begin{eqnarray}
	\Delta t' \pm \sigma_t' &=& \gamma (\Delta t (1-\beta^2)-2  \beta d)  \pm \gamma \sigma_t \;,  \\
	&=& \gamma (\Delta t (1-\beta^2)- 2 \beta d \pm 2 \beta d) \;,  \\
	\Delta t' + \sigma_t' &=& \gamma (\Delta t (1-\beta^2)) \;. \label{eqn:alwayspositive}
\end{eqnarray}
The time difference with a positive uncertainty, Eq.~(\ref{eqn:alwayspositive}), always has a positive value.  The temporal order of the measurements in Bob's reference frame need never flip.

\section{Conclusions}

The uncertainty in time always outruns the time difference induced by the change in reference frames.  Neither Alice nor Bob will ever, with certainty, observe the two measurements swap temporal order.  Of course it is completely possible that when Alice and Bob measure one entangled pair they both get results which are consistent with them both measuring first.  It will only be after many runs of the experiment that the uncertainty becomes evident.  In the experiment proposed above Bob would have to start from the source for each trial and only after many trials will the measurement order, or lack thereof, manifest. 

It may now be said that if a time measurement performed an entangled biphoton is simultaneous in one shared reference frame then it can be considered simultaneous to all measuring observers who do not share a reference frame.  If one were to attempt to determine if the temporal order swaps then an uncertainty will be introduced to make it impossible to determine.  Therefore, there need not be any preferred reference frame for wave function collapse.  The attempt to determine what reference frame the wave function collapse takes place in would lead to an uncertainty that would make it impossible to determine.  That is not to say that all measurements on entangled particles are simultaneous.  There are many situations in which one can determine the order of measurement, but if it can be determined in one shared reference frame then it will be the same or indeterminate in all other reference frames.  

\bibliographystyle{unsrt}
\bibliography{TheBib}

\end{document}